\documentclass{iau} 
\usepackage{graphicx}

\title[IAUS 329.~~Recent advances in non-LTE stellar atmosphere models] %% give here short title %%
{Recent advances in non-LTE stellar atmosphere models}

\author[Andreas Sander]   %% give here short author list %%
{Andreas A.C. Sander$^1$
}

\affiliation{$^1$Institut f{\"u}r Physik \& Astronomie, Universit{\"a}t Potsdam, \\ Karl-Liebknecht-Str. 24-25,
14476 Potsdam, Germany \\ email: {\tt ansander@astro.physik.uni-potsdam.de} %\\[\affilskip]
}

\pubyear{2017}
\volume{329}  %% insert here IAU Symposium No.
\jname{The lives and death-throes of massive stars}
\editors{J.J. Eldridge et al., ed.}
\begin{document}

\maketitle

\begin{abstract}

In the last decades, stellar atmosphere models have become a key tool in
understanding massive stars. Applied for spectroscopic analysis, these models provide
quantitative information on stellar wind properties as well as fundamental
stellar parameters. The intricate non-LTE conditions in stellar winds dictate the development of
adequate sophisticated model atmosphere codes. The increase in both, the computational
power and our understanding of physical processes in stellar atmospheres,
led to an increasing complexity in the models. As a result, codes emerged that can 
tackle a wide range of stellar and wind parameters.

After a brief address of the fundamentals of stellar atmosphere modeling, 
the current stage of clumped and line-blanketed model atmospheres will be discussed.
Finally, the path for the next generation of stellar atmosphere models
will be outlined. Apart from discussing multi-dimensional approaches, I will
emphasize on the coupling of hydrodynamics with a sophisticated
treatment of the radiative transfer. This next generation of models will be able to predict
wind parameters from first principles, which could open new doors for our understanding of 
the various facets of massive star physics, evolution, and death.

\end{abstract}

\firstsection % if your document starts with a section,
\section{Introduction}

A few decades ago, it was Edwin Salpeter who asked Dimitri Mihalas a question
that some might consider to be offensive, namely ``Why in world would anyone want to
study stellar atmospheres?''. A proper answer to this question is nothing less than the justification
of an important keystone in modern astrophysics, and thus D. Mihalas spent several pages on it (\cite[Hubeny \& Mihalas 2014]{HM2014book}). 
Repeating all the arguments from this book would already exceed the page limit of these proceedings,
but essentially the bottom line is that the stellar atmosphere is all we actually see from a star,
and its spectrum is usually the only information we get. 

With the recent advances in asteroseismology and the advent of gravitational wave astronomy this will 
change for some types of objects, but so far clearly not for the majority of stars. Thus, understanding 
the spectrum is the only way to obtain information about them. However, in order to reproduce the spectrum, 
a proper modeling of the stellar atmosphere is necessary. 
Given sufficient observations,
stellar atmospheres can unveil the stellar and wind parameters of a star, such as
$T_\mathrm{eff}$, $\log g$, $L$, $v_\infty$, and $\dot{M}$, provide its chemical
composition, and give insights into its interaction with the environment. Studying stellar 
atmospheres is therefore prerequisite for a plethora of applications and analyses.

\section{Modeling stellar atmospheres}

The atmospheres of hot and massive stars are especially challenging in terms of modeling requirements. 
Their extreme non-LTE situation has to be taken into account and their population numbers can
only be determined by solving the high-dimensional system of equations describing statistical 
equilibrium. Their winds require the calculation of the radiative transfer in an expanding
atmosphere, either by using the so-called ``Sobolev approximation'' (\cite[Sobolev 1960]{S1960book})
or by solving the radiative transfer in the comoving frame (CMF, \cite[Mihalas et al. 1975]{MKH1975}).

For proper modeling, all significant elements in the stellar atmosphere have to be described
by detailed model atoms. This is especially challenging for the elements of the iron group which 
have thousands of levels and millions of line transitions, leading to the so-called ``blanketing'' effect.
Since an explicit treatment is impossible, basically all modern atmosphere codes use a superlevel
concept, going back to \cite[Anderson (1989)]{Anderson1989} and \cite[Dreizler \& Werner (1993)]{DW1993}. 
This is often combined either with some kind of opacity distribution
function (ODF), where the detailed cross sections are not conserved, but resorted to give the correct total 
superlevel cross-section, or an opacity sampling technique where the complex detailed cross sections are
sampled on the frequency grid. 

In an expanding, non-LTE environment, also the determination of the electron temperature stratification is not
trivial. Going back to the ideas of \cite[Uns{\"o}ld (1951, 1955)]{U1951,U1955Book} 
and \cite[Lucy (1964)]{L1964}, temperature corrections can be obtained from the equation of radiative 
equilibrium and it's integral describing the conservation of the total flux. Alternatively, one can also
obtain the corrections from calculating the electron thermal balance going back to \cite[Hummer \& Seaton (1963)]{HS1963}.
Since each of these methods have their strengths and weaknesses, stellar atmosphere codes usually use a combination.

Eventually, the construction of a proper stellar atmosphere model is not just the sum of the 
tasks mentioned in this section. In fact, the tasks are highly coupled, outlining the complexity
of the problem. In CMF radiative transfer, there is an essential coupling in space, while there
is an intrinsic coupling in frequency in the rate equations and temperature corrections. Due to the huge dimensionality
and the non-linear character of the problem, the only way is therefore an iterative algorithm in order to 
establish the consistent solution for all quantities (e.g. radiation field, population numbers, temperature stratification).

Unless the stellar wind is very dense, the vast majority of photons originate in the quasi-hydrostatic 
layers below the supersonic wind. For an easier description, some early models
have treated these two regimes separately, thereby allowing for simplifying approximations 
in each of the domains. However, this ``core-halo'' approach has limits and introduces an artificial boundary. Since
the beginning of the 1990s, the concept of ``unified model atmospheres'' became more common, where the
quasi-hydrostatic and wind regime are described within the same model atmosphere. 

\section{State of the art CMF atmospheres}

The basic scheme for current, state of the art model atmospheres describing stars with stationary winds can be summarized 
as follows: the stellar and wind parameters of a star with an expanding atmosphere are given as input parameters for the
calculation of an atmosphere model. After assuming a certain starting approach, the equations of statistical
equilibrium and radiative transfer are iteratively solved where such iteration must be ``accelerated'' by a suitable algorithm \cite[(e.g. Hamann 1987)]{H1987}.
In addition, the temperature stratification has to be updated in parallel.
When the total changes in this iterative scheme drop below some prespecified level, the
atmosphere model is considered to be converged. In a last step, the emergent spectrum (in the observer's frame)
is calculated based on the converged model atmosphere.

Regarding the techniques, current, state of the art model atmospheres share the following properties:
the models can account for many elements, including the iron group. There is no artificial boundary between the 
subsonic (quasi-hydrostatic) and supersonic 
(wind) regime. For the latter, a predescribed velocity law $v(r)$ is adopted while a 
more detailed treatment is applied to the quasi-hydrostatic part. Density inhomogeneities (aka ``clumping'') can be tackled in an 
approximate way. For more details see Gr{\"a}fener (2017, these proceedings). For the detailed radiative transfer,
typically either a Monte Carlo (MC) or a comoving frame (CMF) approach is implemented.

In the CMF approach, the radiative acceleration $a_\mathrm{rad}$ is obtained by an evaluation
of the integral
\begin{equation}
  a_\mathrm{rad}(r) = \frac{4\pi}{c} \frac{1}{\rho(r)} \int\limits_{0}^{\infty} \kappa_\nu(r) H_\nu(r) \mathrm{d}\nu
\end{equation}
in the comoving frame. While approximate treatments following CAK \cite[(named after Castor, Abbott, \& Klein 1975)]{CAK1975} 
or it's later extensions have their strength in reducing the description of a $a_\mathrm{rad}$ into a (semi\hbox{-})analytical form, allowing a
fast calculation, they neglect effects like multiple scattering and essentially break down for dense winds.
In contrast, the CMF calculation implicitly includes various effects and thus works for all line-driven
winds, ranging from classical OB stars to Luminous Blue Variables (LBVs) and Wolf-Rayet (WR) stars. Even low-mass stars with line-driven winds,
such as O subdwarfs or WR-type central stars of planetary nebulae, can be calculated in this way.

\section{Recent advances in state of the art atmospheres}

The current generation of stellar atmosphere models has reached high complexity. Updating or
extending the model codes is therefore often a non-trivial task. Apart from the new features
eventually visible to the general user, a lot of technical aspects have to be taken into account and
the developers have to check whether the original premises made for their code
still hold in all of the current (and intended) applications. This includes
technical and physical aspects, such as start approximations, boundary conditions, blanketing treatment, the
description of $v(r)$ in the subsonic domain, microturbulence, and clumping, the calculation of the emergent spectrum, or the accuracy and
completeness of the atomic data. Some of these tasks are implemented at a fundamental level of the code and adjusting
them can therefore result in a significant amount of work, often not or just partly visible to the general audience.
In order to shed a bit more light on the great amount of work that has been performed in this field on all scales
during approximately the last decade, the following list of advances does not only cover such updates which are 
immediately interesting for the user, but also some more technical aspects as far as they were documented. The
codes are listed in alphabetical order:

{\underline{\it CMFGEN}} (\cite[Hillier 1987, 1990, 1991; Hillier \& Miller 1998]{H1987,H1990,H1991a,HM1998}; diagnostic: UV, optical, IR) 
uses a detailed CMF radiative transfer, fully accounting for line blanketing. It
received a major extension in the last decade by the inclusion of the time-dependent terms in the rate equations and the addition of
time-dependent radiative transfer modules in order to allow for the calculation of supernovae spectra \cite[(Dessart \& Hillier 2010)]{DH2010}. 
Regarding stars, the concept of ``hydrostatic iterations'' was introduced (\cite[Martins \& Hillier 2012]{MH2012})
where the velocity description is updated a few times in order to better match with the hydrostatic equation in the
subsonic part. For the emergent spectrum, rotational broadening can now be considered (\cite[Hillier et al. 2012]{Hillier+2012}) and the code can
handle depth-dependent Doppler and Stark broadening. In 2016, the option to handle H$^{-}$ opacity has
been added. To speed up the calculations of a single model, CMFGEN can also make use of more than one core, using
partial code parallelization.

{\underline{\it FASTWIND}} (\cite[Santolaya-Rey et al. 1997; Puls et al. 2005]{SPH1997,Puls+2005}; diagnostic: optical, IR)
focusses on calculation speed and therefore performs a split into so-called ``explicit'' and ``background'' elements.
Originally, only the explicit elements are treated with a CMF approach, while background elements are tackled with Sobolev.
The line blanketing is approximated, leading to an additional performance gain. The more recent developments of FASTWIND 
are described in \cite[Rivero González et al. (2011,\,2012)]{RiveroGonzalez+2011,RiveroGonzalez+2012}. The photospheric line
acceleration is now properly treated and, in contrast to earlier
versions, important background lines are now also treated in the CMF. A new version that is able to treat all lines in the
CMF and thus also extends the diagnostic range to the UV is in development and described by Puls (2017, these proceedings).
Furthermore, FASTWIND is now also able to consider X-ray emission from wind-embedded shocks \cite[(Carneiro et al. 2016)]{Carneiro+2016}.

{\underline{\it Krti{\v c}ka \& Kubat}} (\cite[Krti{\v c}ka \& Kubat 2004,\,2009; Krti{\v c}ka 2006]{KK2004,KK2009,K2006}; no emergent spectra) have
a code that is focussed on predicting mass-loss rates by solving the
hydrodynamical equation together with the rest of the atmosphere iteration. While their code originally relied on the Sobolev 
approximation, the line force is now also calculated in the
CMF \cite[(Krti{\v c}ka \& Kubat 2010)]{KK2010}. Their calculations can furthermore account for turbulent broadening of the line profiles.

{\underline{\it PHOENIX}} (\cite[Hauschildt 1992; Hauschildt \& Baron 1999,\,2004]{H1992,HB1999,HB2004}; diagnostic: UV, optical, IR) is a CMF-based code,
commonly applied in the modeling of cool stars and supernovae, but in principle also suitable for hot stars (see, e.g. \cite[Hauschildt 1992]{H1992}). 
So far unmet by other codes, the developers of PHOENIX have started with a 3D branch of their code in addition to their
standard 1D branch (see Sect.\,\ref{sec:multid}). Some 3D test results also help to improve their 1D branch, e.g. in the case of obtaining limb darkening coefficients.

{\underline{\it PoWR}} (\cite[Hamann 1985, 1986; Hamann \& Schmutz 1987; Koesterke et al. 2002; Gr{\"a}fener et al. 2002]{H1985,H1986,HS1987,KHG2002,GKH2002}; diagnostic: UV, optical, IR) 
also uses the CMF radiative transfer and fully accounts for line blanketing. Originally developed for Wolf-Rayet stars, it has since been significantly extended to be nowadays 
applicable to any hot star. For an accurate photospheric density stratification, the quasi-hydrostatic domain is now treated self-consistently, i.e. the velocity
and density stratification are constantly updated in the course of the iteration \cite[(Sander et al. 2015)]{Sander+2015}. For a proper calculation
of the emergent spectrum, the formal integral accounts for line broadening, including the linear and quadratic Stark effect, 
rotation \cite[(Shenar et al. 2014)]{Shenar+2014}, and mircoturbulent broadening. On the more technical side, the blanketing treatment has been updated
and the superlevels now strictly separate the different parities. 
Furthermore, the temperature correction can now alternatively be obtained 
via thermal balance instead of radiative equilibrium. 

In order to predict mass-loss rates, the PoWR code has also a recently added branch to calculate 
hydrodynamically consistent models (see also Sect.\,\ref{sec:hydro}.) Although based on the ideas of the first efforts from \cite[Gr{\"a}fener \& Hamann (2005, 2008)]{GH2005,GH2008} for Wolf-Rayet models, the technical details of the new method differ significantly in detail.

{\underline{\it WM-basic}} (\cite[Pauldrach 1987, Pauldrach et al. 1994,\,2001]{P1987,Pauldrach+1994,PHL2001}; diagnostic: UV, optical) applies the
Sobolev approximation for the line transfer. It can account for an EUV and X-ray shock source function and is able to calculate hydrodynamically
consistent models. In the past decade, WM-basic has also been extended to 
calculates supernova ``snapshot'' spectra \cite[(e.g. Pauldrach et al. 2012)]{Pauldrach+2012}. Furthermore, 
with the inclusion of stark broadening \cite[(Kaschinski et al. 2012, Pauldrach et al. 2014)]{Kaschinski+2012,Pauldrach+2014}, the diagnostic range 
was extended to the optical regime.

An alternative to CMF-based codes are {\underline{\it Monte Carlo codes}}, where the radiative acceleration is obtained
by following energy or photon packages throughout the atmosphere. Basically all current approaches have built up on the concepts of \cite{FA1986}. 
Using so-called ``moving reversing layers'', mass-fluxes are obtained as eigenvalues by \cite{L2007a}. While already outlined 
in \cite{LS1970}, it requires the modern generation of computers to calculate a
larger range of models and study details like the influence of the microturbulence, metallicity or the potential roots 
of the so-called ``weak-wind problem''. Consequently, this approach was not just improved in technical details, such
as improving the lower boundary condition or updating the line list, but 
foremost applied to various parameter ranges \cite[(Lucy 2010a,\,2010b,\,2012)]{L2010a,L2010b,L2012}. A different 
approach using MC techniques was taken by \cite{MV2008}, who found solutions for the velocity
field with the help of the Lambert~W-function by adopting a semi-analytic description for the radiative acceleration in the
(isothermal) wind. This allows to achieve local hydrodynamical consistency in their models, in contrast to the 
earlier mass-loss predictions from \cite[Vink et al. (1999,\,2000,\,2001)]{VdKL1999,VdKL2000,VdKL2001} where only a global consistency was
guaranteed. The new approach was extended to 2D and applied to rotating massive O stars in \cite{MV2014},
allowing to study effects like oblateness and gravity darkening. A new MC-based code is also currently in development by \cite{NS2015}.

\section{What's next?}

When discussing the potential next steps for non-LTE stellar atmosphere models, things such as 2D/3D calculations, 
hydrodynamical consistency, consistent X-ray treatment, multi-component 
winds or non-monotonic velocity fields are coming to mind.  For the sake of time (or page) limitation, 
only the first two will be discussed below. But apart from this ``wishlist'' from the user's point 
of view one has also to be aware of the more imperceptive challenges in the development process which only
become visible if we take the developer's point of view. To identify 
and tackle ``problematic'' parameter regimes in the codes is already a
task in itself. Finding ``good'' compromises between accuracy, numerics, and computational performance is also a constant
topic for code developers as we see more and more applications coming up where whole grids of models are required
and manual checks for each model become practically impossible. On top of these 
technical aspects, there are scientific challenges like a better description of turbulence and clumping. 
And of course there
is one of the biggest underlying challenges introducing unknown uncertainties basically in any model, namely
the atomic data. This is not just a question of completeness -- also more unapparent aspects like the details of 
superlevel approximations or the handling of ionization cross-sections can have an effect on the results. 
While finding good constraints on these is one important future task, it might be even more important, 
especially for the user, to simply keep in mind that there are systematic errors and simplifying assumptions and
thus one should not overestimate the precision of the derived spectra.

\subsection{2D and 3D approaches}
  \label{sec:multid}

Given the significant computational effort when using a CMF radiative transfer, much more work 
in the field of multi-D approaches, which are necessary for studying non-spherical stars or structures,
has been done with Monte Carlo models. 
As mentioned in the previous section, \cite{MV2014} modeled an axially
symmetric rotating star and found an equatorial decrease in the mass-loss, implying that 
the total mass-loss rate would be lower than in the 1D-spherical case, thereby contradicting the
predictions from \cite{MM2000}.
On the other hand, there is the very sophisticated and ambitious idea of a full 3D CMF 
radiative transfer. This has been started by \cite{HB2006} using their PHOENIX code, now 
termed PHOENIX/3D for this branch. In a series of currently 11 papers, they present the ideas and
methods of their 3D branch and show several test calculations where they compare 3D results to their
1D counterparts. While their efforts and progress is impressive, it also shows that detailed 3D non-LTE
atmospheres will not become a standard tool in the near future, since the current calculations require
large supercomputers as soon as complex ions are used. In \cite{HB2014}, they presented a small test
case using only $62$ non-LTE levels and a very large case using $4686$ non-LTE levels. For a single
iteration, i.e. one the solution of the 3D radiative transfer plus one solution of the statistical
equations etc., already the small case requires about half a year on a single core using an Intel 
Xeon E5420 CPU with $2.50$ GHz clock-speed. For the large, but much more realistic case this value 
rises to $4300$ years for a single iteration, thus leading to the effect that \cite{HB2014} decide to
give the total linear calculation time, i.e. CPU time until model convergence, as $15\,\mu$Hubble or
roughly $215\,000\,$years. While the corresponding wall clock times could be lowered significantly by
parallelization, the number of cores required to bring this down to the order of days would be enormous.

\subsection{Hydrodynamically self-consistent atmosphere models}
  \label{sec:hydro}

While nowadays stellar atmosphere models are mostly used to measure the wind parameters ($v(r), \dot{M}$) of
a star, a new generation of models is designed to predict these.
The key ingredient for this task is the inclusion of the hydrodynamic (HD) equation
\begin{equation}
  \label{eq:hydro}
     v \left( 1 - \frac{a_\mathrm{s}^2}{v^2} \right) \frac{\mathrm{d} v}{\mathrm{d} r} = a_\mathrm{rad}(r) - g(r) + 2 \frac{a_\mathrm{s}^2}{r} - \frac{\mathrm{d} a_\mathrm{s}^2}{\mathrm{d} r}\mathrm{,}
\end{equation}
which has to be fulfilled at all depths in a self-consistent atmosphere model. The predictablity of the 
wind parameters is achieved due to the
additional constraint introduced by the hydrodynamic Eq.\,(\ref{eq:hydro}) and its critical point. 
The requirement to have a smooth transition of $v(r)$ through the critical point can be translated into a 
condition for the mass-loss rate $\dot{M}$. Thereby the model predicts this fundamental wind parameter 
purely from the given stellar parameters.

\begin{figure}[b]
\begin{center}
 \includegraphics[width=0.7\textwidth]{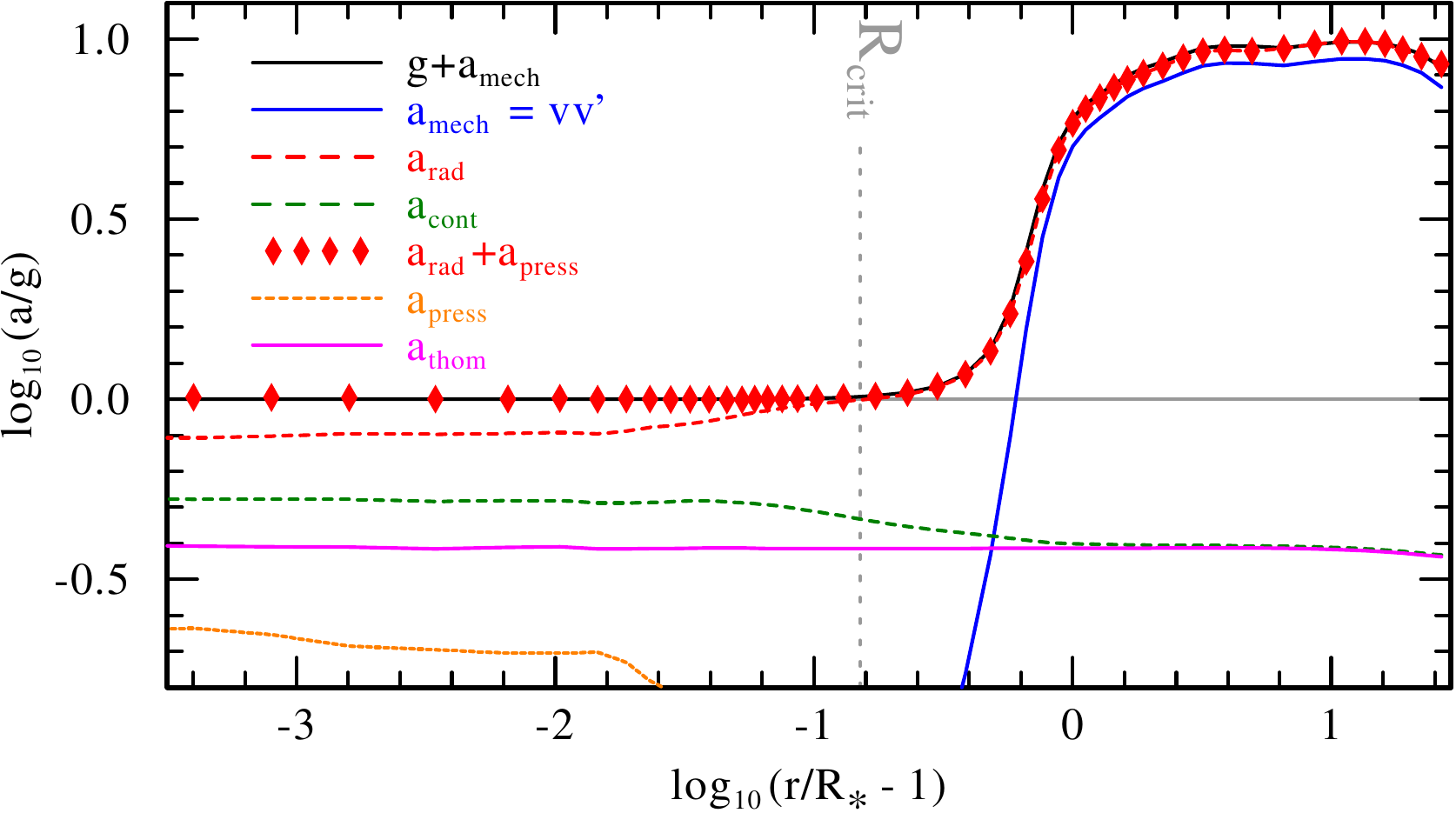} 
 \caption{Acceleration stratification for an O supergiant model: The red diamonds ($a_\mathrm{rad} + a_\mathrm{press}$) match 
          the black curve ($g + a_\mathrm{mech}$), illustrating that the HD Eq.\,(\ref{eq:hydro}) is fulfilled.}
   \label{fig:zpuphydro}
\end{center}
\end{figure}
While the idea is relatively simple and goes back to \cite{LS1970}, its actual implementation is not.
Early efforts using a pure CMF line force implementation were made by \cite{PPK1986} and a 
Sobolev-based implementation became part of WM-basic with \cite{PHL2001}. The concept is also applied
in the more theoretical works of \cite{KK2004}, first via Sobolev and later in \cite{KK2010} with 
a CMF approach. The first complete implementation into a CMF-based analysis code was done by
\cite{GH2005} using the PoWR code. Their implementation technique, based on a generalized force multiplier 
concept, was successfully applied to a WC and later also to a grid of WN models \cite[(Gr{\"a}fener \& Hamann 2008)]{GH2008}.
A new implementation utilizing a different technique was recently added to the PoWR code \cite[(Sander et al. 2015b, 2017 submitted)]{Sander+2015b},
finally allowing to also calculate hydrodynamically consistent models for OB stars. An example for an O supergiant
model based on $\zeta\,$Pup can be seen in Fig.\,\ref{fig:zpuphydro}. Recently, J. Puls and J. Sundqvist have
also started to implement consistent hydrodynamics into their new version of FASTWIND.

\section{Summary}

Compared to earlier decades, the current generation of non-LTE expanding stellar atmosphere models has a
significantly increased applicability, ranging from OB stars via transition stages like LBVs or Of/WN stars
up to the classical Wolf-Rayets. Moreover, they are applicable for hot low-mass stars, such
as central stars of planetary nebulae. With more codes allowing for a such a broad range of applications, more
benchmarking will be possible, allowing to cross-check results between different codes and methods and identify
problems. PoWR, CMFGEN, and FASTWIND can also treat X-rays from wind-embedded 
shocks, thereby extending the diagnostic range into this wavelength regime.

A completely different branch has been opened by CMFGEN, PHOENIX, and WM-basic with their option to calculate so-called
supernova ``snapshot'' spectra. By adding the time-dependent terms in the rate equations and the radiative transfer,
atmosphere codes can be used to analyze supernova spectra and study their evolution.

In the years to come, hydrodynamically consistent models will provide a new generation of model atmospheres which
can predict wind parameters from a given set of stellar parameters, thereby opening up
a third branch next to MC and CAK-like techniques. Due to the local consistence of the models and the possibility
to calculate emergent spectra, the results can immediately be cross-checked with observations. Apart from a Sobolev-based 
implementation in WM-basic at the beginning of the century, HD-consistent models can now also be calculated with PoWR and potentially with FASTWIND.

Although mainly used for cool stars, significant efforts to obtain a 3D CMF radiative transfer have been made
with PHOENIX/3D. Unfortunately, massive parallelization is required
to reach manageable wall clock times for 3D models, and thus 1D model atmospheres will 
remain the standard tool for massive stars in the near future.

\end{document}